\documentclass[modern]{aastex61}

\usepackage{graphicx}
\usepackage{amsmath}
\usepackage{amssymb}
\usepackage{enumitem}

\newcommand\mybar{\kern1pt\rule[-\dp\strutbox]{.8pt}{\baselineskip}\kern1pt}

\setlist[itemize]{noitemsep, topsep=0pt, leftmargin=*}

\shorttitle{Relativistic Effects in Milky Way Disk}
\shortauthors{Loeb}

\begin{document}

\title{Spurious Radial Migration from Relativistic Effects in the
  Milky-Way Disk}

\author{Abraham Loeb}
\affiliation{Astronomy Department, Harvard University, 60 Garden
  St., Cambridge, MA 02138, USA}

\begin{abstract}
The gradient of the gravitational redshift in the potential of the
Milky-Way induces an apparent spurious radial migration.  I show that
this effect is simply related to the local acceleration, which was
measured recently by {\it Gaia} eDR3, implying a spectroscopic shift
of $-2.4\times 10^{-2}(r/8~{\rm kpc})^{-1}~{\rm
  km~s^{-1}~kpc^{-1}}$. The transverse Doppler effect yields a
comparable contribution. The spurious radial velocity from both
relativistic effects amounts to crossing a major portion of the
Milky-Way disk during the age of the universe, and must be corrected
for in any future measurement of the actual radial migration of stars.
\end{abstract}

\section{Introduction}

According to General Relativity, time is dilated in a gravitational
potential well, $\phi$, relative to a distant observer.  In the weak
field regime, the shift in the ticking rate of a clock leads to a
spectroscopic velocity offset in the radial $r$ direction away from
the observer \citep{1972gcpa.book.....W},
\begin{equation}
{{\Delta v_r}\over c}=-{\Delta \phi\over c^2} .
\label{one}
\end{equation}
The related gravitational redshift was confirmed recently on a
millimeter scale using atomic clocks \citep{2022Natur.602..420B}, and
was detected previously in spectral lines emitted from compact stars,
such as white dwarfs
\citep{1967AJ.....72Q.301G,2005MNRAS.362.1134B,2012ApJ...757..116F} or
neutron stars \citep{2002Natur.420...51C}.

The latest data from {\it Gaia} eDR3 implied that the Milky Way has a
nearly flat rotation curve with a local circular speed $v_c\approx
230~{\rm km~s^{-1}}$ that declines gently outwards by an amount
$(dv_c/dr)=-(1.7\pm 0.1)~{\rm km~s^{-1}~kpc^{-1}}$
\citep{2019ApJ...871..120E,2019ApJ...870L..10M}.

\section{Results}

Equation (\ref{one}) implies an apparent spurious radial velocity with
a radial gradient,
\begin{equation}
{\partial v_r\over \partial r}=- {1\over c}{\partial \phi\over
  \partial r} = - {1\over c}{v_c^2\over r} ,
\label{two}
\end{equation}
where $g=-(\partial \phi/\partial r)=-(v_c^2/r)$ is the local
gravitational acceleration. This leads to the spurious inference of
radial recession (net redshift) away from the Sun for stars closer in to
the Galactic center and radial approach (net blueshift) of stars
farther out than the Sun's Galactocentric distance.

Data from {\it Gaia} eDR3 \citep{2021A&A...649A...9G} implies
$g=-(2.32\pm0.16)\times 10^{-8}~{\rm cm~s^{-2}}$, yielding a velocity
gradient,
\begin{equation}
{\partial v_r\over \partial r}= - 2.4\times 10^{-2}\left({r\over
  8~{\rm kpc}}\right)^{-1}~{\rm km~s^{-1}~kpc^{-1}}.
\label{three}
\end{equation}
The spurious radial velocity is significant, as it amounts to crossing
a major portion of the Milky-Way disk during the age of the universe.
An actual radial migration of this magnitude is inferred by other
means \citep{2018ApJ...865...96F,2022MNRAS.511.5639L}.  The velocity
amplitude is smaller than current measurement errors
\citep{2019ApJ...871..120E,2019ApJ...870L..10M}.

Time dilation is also sourced by the Lorentz factor, leading to the
transverse Doppler effect, which is not accounted for in
non-relativistic calculations. At small velocities, the effect is
second order in the velocity over the speed of light.  Owing to the
virial theorem, $\langle (v/c)^2 \rangle = -(1/2)\langle
(\phi/c^2)\rangle$, the transverse Doppler effect is comparable in
magnitude to the gravitational redshift effect, and leads to a
spurious radial velocity gradient,
\begin{equation}
{\partial v_r\over \partial r}= {1\over 2}c {\partial \over \partial
  r} \left({v_c\over c}\right)^2 = 9\times 10^{-2}~{\rm
  km~s^{-1}}{\partial\over \partial r} \left({v_c\over 230~{\rm
    km~s^{-1}}}\right)^2 .
\label{four}
\end{equation}

\section{Implications}

The sum of the relativistic effects in equations~(\ref{three}) and
(\ref{four}) implies a spurious radial migration, that amounts to
crossing a major portion of the Milky-Way disk during the age of the
universe. An actual migration of this magnitude is inferred by other
means \citep{2018ApJ...865...96F,2022MNRAS.511.5639L}, and its proper
direct measurement in the future would need to correct for the
relativistic effects mentioned here.

\bigskip
\bigskip
\section*{Acknowledgements}

This work was supported in part by Harvard's {\it Black Hole
  Initiative}, which is funded by grants from JFT and GBMF. 

\bigskip
\bigskip
\bigskip

\bibliographystyle{aasjournal}
\bibliography{m}
\label{lastpage}
\end{document}